\documentclass[aps,prl,showpacs,floatfix,twocolumn,superscriptaddress]{revtex4}
\usepackage{microtype}
\usepackage{amssymb}
\usepackage{amsmath}
\usepackage{graphicx}
\usepackage[latin1]{inputenc}
\begin{document}
\setlength{\parindent}{0pt} \newcommand{\llangle}{\langle \langle}
\newcommand{\rrangle}{\rangle \rangle}
\title{Measuring the momentum of a nanomechanical oscillator through the
use of two tunnel junctions}
\author{C.B. Doiron}
\affiliation{Department of Physics, University of Basel,
CH-4056 Basel, Switzerland}
\author{B. Trauzettel}
\affiliation{Institute of Theoretical Physics and Astrophysics, University of
W{\"u}rzburg, D-97074 W{\"u}rzburg, Germany}\affiliation{Department of Physics, University of Basel,
CH-4056 Basel, Switzerland}
\author{C. Bruder}
\affiliation{Department of Physics, University of Basel,
CH-4056 Basel, Switzerland}


\begin{abstract}
We propose a way to measure the momentum $p$ of a nanomechanical oscillator.
The $p$-detector is based on two tunnel junctions in an
Aharonov-Bohm-type setup. 
One of the tunneling amplitudes depends on the motion of the
oscillator, the other one not. Although the coupling between
the detector and the oscillator is assumed to be linear in the position $x$ of
the oscillator, it turns out that the finite-frequency
noise output of the detector will in general 
contain a term proportional to the
momentum spectrum of the oscillator. This is a true quantum phenomenon,
which can be realized in practice
if the phase of the tunneling amplitude of the detector is tuned by the
Aharonov-Bohm flux $\Phi$ to a $p$-sensitive value.
\end{abstract}

\date{November 2007} \maketitle

In nano-electromechanical (NEM) systems, it is the position of the
oscillator that typical measurement devices (like tunnel junctions or
single electron transistors) are coupled to. Using these detectors,
position measurements with sensitivities close to the standard quantum
limit have already been observed
\cite{knobel2003,lahaye2004,naik2006}. From a fundamental point of
view, it is desirable to go further, i.e.~to prepare and manipulate
NEM oscillators in the quantum regime. A quantum NEM system would
allow us to study an ideal realization of a continuous variable
quantum system \cite{braunstein}. The exploration of such systems has
to be seen as complementary to the wide study of two-level systems
done in the context of quantum computing.

\begin{figure*}[htb]
\begin{center}
\leavevmode \includegraphics[width=14cm]{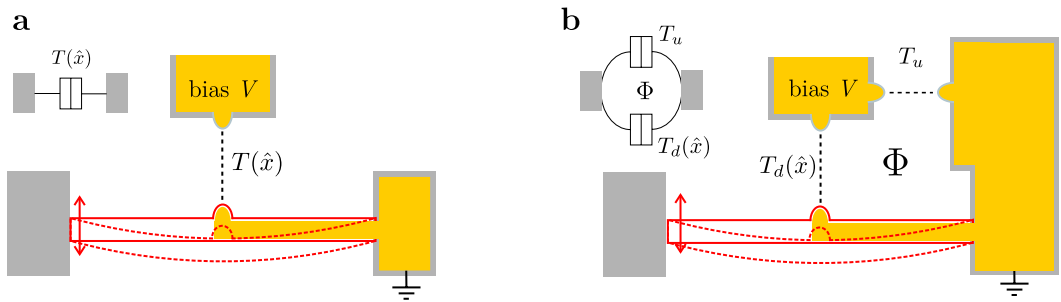}
\caption{(color online) {\bf a} {\it Position detector.} The
figure shows schematically a position detector of the motion of a
NEM oscillator (red). The detector is based on a tunnel junction
with a tunnel matrix element $T(\hat{x})$ which depends 
on the position of the oscillator. 
The shaded regions (yellow) are assumed to be conducting. {\bf b} {\it
Position and/or momentum detector.} The figure illustrates a
detector which contains two tunnel junctions that form a loop
threaded by a magnetic flux $\Phi$. Tuning the flux will 
change the performance of the detector from being able to
detect the power spectrum of the {\it position operator} of the
oscillator to being able to detect the power spectrum of the {\it
momentum operator} of the oscillator. For clarity, the two insets
show a simplified illustration of the detectors in {\bf a} and
{\bf b}.} \label{fig_setup}
\end{center}
\end{figure*}

In order to be able to fully characterize a continuous variable
quantum system that is described by two non-commuting operators
$\hat{x}$ and $\hat{p}$, we need to be able to measure expectation
values of moments of both of them \cite{duan2000}. Only this allows,
for instance, to detect the entanglement between two (or more) NEM
devices \cite{eisert2004}. The literature already contains proposals
regarding quantum measurements of the momentum of macroscopic objects
like those used for gravity-wave detection \cite{momentumgw}. However,
none of these proposals have been realized in practice. In this
Letter, we propose a realistic and feasible way to measure the
momentum of a \emph{nanometer-sized} resonator. This is a non-trivial
task since the coupling between the detector and the oscillator is
naturally described by an $x$-dependence but not a $p$-dependence.
Nevertheless, the proposed setup (shown in Fig.~\ref{fig_setup}b)
allows for a measurement of the momentum spectrum $S_p(\omega) = \int
dt e^{i \omega t} \langle \{ \hat{p}(t), \hat{p}(0) \} \rangle$ of the
oscillator. This can be done because we have found a way to tune the
phase of the tunnel coupling term that is sensitive to the position of
the oscillator by an Aharonov-Bohm (AB) flux $\Phi$, see
Fig.~\ref{fig_setup}b. Related setups have been investigated recently
in the context of dephasing due to the coupling of an AB ring
structure to a NEM device \cite{armour2001}.
%
%
A typical position detector which has been analyzed theoretically
in great detail
\cite{bocko1988,yurke1990,schwabe1995,mozyrsky2002,clerk2004b,clerk2004a,wabnig2005}
and experimentally realized
\cite{cleland2002,flowers-jacobs2007} is depicted in
Fig.~\ref{fig_setup}a. It shows a single tunnel junction coupled
to a NEM oscillator. A thorough analysis of the coupled quantum
system leads to the result that the output
signal of the detector is sensitive to the position spectrum
$S_x(\omega) = \int dt e^{i \omega t} \langle \{ \hat{x}(t),
\hat{x}(0) \} \rangle$ of the oscillator. The modification of the
detector shown in Fig.~\ref{fig_setup}b instead allows for a
measurement of $S_p(\omega)$.

The Hamiltonian of the coupled system $H=H_{\rm osc}+H_B+H_{\rm tun}$
is the sum of the Hamiltonian of the (quantum) harmonic oscillator
$H_{\rm osc}$ (with mass $M$ and frequency $\Omega$), 
the bath Hamiltonian $H_B$ (describing the leads of the
detector),
and the tunneling Hamiltonian $H_{\rm tun}$ (which couples the dynamics of the
electrons that tunnel across the junction
to the motion of the oscillator):
\begin{align}
H_{\rm osc} &= \hbar \Omega (a^\dagger a + \frac{1}{2}) =
\frac{\hat{p}^2}{2M} + \frac{M \Omega^2 \hat{x}^2}{2} \; ,\\
H_{B} &= \sum_k \varepsilon_k c^\dagger _k c_k + \sum_q \varepsilon_q c^\dagger _q c_q \;, \\
H_{\rm tun} &= T(\hat{x}) Y^\dagger \sum _{k,q}c^\dagger _k c_q +
T^\dagger (\hat{x}) Y \sum _{k,q}c^\dagger _q c_k \;.
\end{align}
Here, $k$ $(q)$ is a wave-vector in the right (left) lead,
$c^{(\dagger)}$ is the electron annihilation (creation) operator, and
$Y^{(\dagger)}$ is an operator that decreases (increases) $m$, the
number of electrons that have tunneled through the system, by one. It
allows one to keep track of the transport processes during the
evolution of the system. For typical nanoresonators, the mass $M$
varies between $10^{-18}$ and $10^{-15}$ kg, and the resonance
frequency is usually $1 \mathrm{MHz} < \Omega / 2\pi < 1 \mathrm{GHz}$
\cite{ekincirsi}.

We first discuss the model in the standard configuration shown in
Fig.~\ref{fig_setup}a and later on describe 
the 
new setup in Fig.~\ref{fig_setup}b. For
small displacements with respect to the tunneling length (which is
the relevant regime in typical experiments on NEM devices), the
tunneling amplitude $T(\hat{x})$ can be taken as a linear function
of $\hat{x}$, namely $T(\hat{x}) = (e^{i\varphi_0}/2\pi \Lambda) 
\left( \tau_0 + e^{i \eta}\tau_1 \hat{x} \right)$,
where $\tau_0$ and $\tau_1$ are real, and $\Lambda$ is the density of
states. The phases $\varphi_0$ and
$\eta$ describe details of the detector-oscillator coupling
\cite{foot3}. 
It can be shown that our detector is
quantum-limited for any $\eta$, according to the definition of a
quantum-limited detector in Ref.~\cite{clerk2004a}.
Within the single-junction setup
(Fig.~\ref{fig_setup}a), the relative phase $\eta$ is sample-dependent
and cannot be tuned experimentally. In a typical device, the
$x$-dependence of the phase of $T(\hat{x})$ is much weaker than the
$x$-dependence of the amplitude. Then, we can set $\eta \simeq 0$, and
the tunnel junction acts as a position-to-current amplifier where the
frequency-dependent current noise $S_I (\omega)$ of the detector
contains a term proportional to the position spectrum $S_x(\omega)$ of
the oscillator, i.e. $\Delta S_I (\omega) = S_I(\omega) - 2e\langle I
\rangle \approx \lambda_x^2 S_x (\omega)$ with $\lambda_x$ the
gain of the amplifier \cite{clerk2004b,doiron2007}.

We now demonstrate that a tunnel junction with a 
phase $\eta = \pi/2 \,
{\rm mod} \, \pi$ acts as a momentum detector and
$\Delta S_I(\omega) \approx \lambda_p^2 S_p
(\omega)$, where $\lambda_p$ is the gain of the
momentum-to-current linear amplifier. The critical requirement to
build a momentum detector is to be able to vary $\eta$
experimentally. This can be done using the AB-type setup shown
in Fig.~\ref{fig_setup}b:
a metallic ring
where one arm is a standard tunnel junction position detector with
tunneling amplitude $T_d(\hat{x})$, and the other arm is a
position-independent tunnel junction with tunneling amplitude
$T_u$ \cite{footMZ}. The total transmission amplitude $T(\hat{x},\Phi)$
of the device is the sum of both tunneling amplitudes \cite{foot4}. 
Since 
only one arm shows a position-dependence, the
induced phase difference between the two arms affects
the position-independent and the position-dependent
parts of the tunneling amplitudes $\tau_0$ and $\tau_1$
in a different way.
Explicit calculation shows that the tunneling amplitude is
given (up to a global gauge-dependent phase factor) by
$T(\hat{x},\Phi)= ( \tau_0 (\Phi) + e^{i
  \eta(\Phi)} \tau_1 \hat{x}) / (2 \pi \Lambda)$ with
\begin{eqnarray} \label{tau0p2}
\tau^2_0 (\Phi) &=& \tau_{0,d}^2  +  \tau_{0,u}^2 + 2 \tau_{0,d} \tau_{0,u} 
\cos \Bigl( 2\pi \frac{\Phi}{\Phi_0} + \varphi_{0,d}-\varphi_{0,u} \Bigr),
\nonumber \\
\eta (\Phi) &=&
2\pi \frac{\Phi}{\Phi_0} + \varphi_{1,d}-\varphi_{0,u} \nonumber \\
&-& \mathrm{Arg} \Bigl( \tau_{0,u} + e^{i (2\pi
  \frac{\Phi}{\Phi_0} + \varphi_{0,d}-\varphi_{0,u}) }  \tau_{0,d} \Bigr),
\end{eqnarray}
where we have defined $T_u \equiv e^{i \varphi_0,u} \tau_{0,u}$, $T_d
\equiv e^{i \varphi_0,d} \tau_{0,d} + e^{i \varphi_{1,d}} \tau_{1,d} \hat{x}$,
$\tau_1 \equiv \tau_{1,d}$, and $\Phi_0 = h/e$ \cite{foot5}.
The position-independent part of the tunneling amplitude
$\tau_0(\Phi)$ displays the standard AB oscillations as a function of
flux. Likewise, the relative phase $\eta(\Phi)$ shows a distinct
dependence on the flux. Importantly, for $\tau_{0,u} > \tau_{0,d} $,
the phase $\eta(\Phi)$ can be tuned continuously in the whole range
$[-\pi,\pi]$. In the limit, where $\tau_{0,u} \gg \tau_{0,d}$, $\eta
(\Phi) \sim 2\pi \frac{\Phi}{\Phi_0}+ \eta (\Phi=0)$ varies linearly
with the applied flux. In the opposite regime $\tau_{0,u} \ll
\tau_{0,d}$, $\eta$ no longer depends on $\Phi$. Therefore, it is
crucial to put the tunneling amplitudes in the regime where
$\eta(\Phi)$ can be tuned to $\pi/2$. We will show below that a
feasible way to calibrate $\eta(\Phi)$ to the $p$-sensitive point
$\pi/2$ is a measurement of the flux dependence of the current through
the AB detector.

Similarly to Refs.~\cite{clerk2004b,doiron2007}, we study the coupled system
using the quantum equation of motion for the charge-resolved density matrix
within the Born-Markov approximation, assuming that  $eV \gg \hbar \Omega$.

It has been derived previously
that, under the assumption that the tunneling
amplitude depends linearly on $\hat{x}$, the equation of motion for the
reduced density matrix of the oscillator is of Caldeira-Leggett
form \cite{mozyrsky2002,clerk2004b,wabnig2005}. Thus,
it contains both a damping and a
diffusion term. When the electron temperature is much smaller than the applied
bias (and taking $V>0$), the detector-induced
damping coefficient is $\gamma_+ = \hbar \tau_1^2 / (4\pi M)$ and the
diffusion coefficient is $D_+=2M \gamma_+ k_B T_{\mathrm{eff}}$ with
$T_{\mathrm{eff}} = eV/2k_B$.

In general, the oscillator is not only coupled to the detector but
also to the environment.  The coupling to this additional bath is
controlled via $\gamma_0 = \Omega/ Q_0$ (related to the finite quality
factor $Q_0$ of the mode, which in current experiments varies from
$10^3$ to $10^6$ \cite{ekincirsi}) and the associated diffusion
constant $D_0 = 2 M \gamma_0 k_B T_{\mathrm{env}}$ that must be added
to the detector-induced damping and diffusion constants to find the
total damping coefficient $\gamma_{\mathrm{tot}}=\gamma_+ + \gamma_0$
and the total diffusion coefficient
$D_{\mathrm{tot}}=D_0+D_+$. $T_{\mathrm{env}}$ denotes the temperature
of the environment. In typical experiments, it varies from 30mK to
10K. Within our model, all the these system parameters are independent
of the applied flux.

It is now straightforward to calculate the current and the
current noise of the detector. We skip the details here 
(see Ref.~\cite{doiron2007} for $\eta=0$) and
directly turn to the results. The average
current of the detector is given by
\begin{eqnarray} \label{idet}
I &=&  \frac{e^2 V}{h} \left( \tau_0^2 + 2 \cos \eta \tau_0 \tau_1 \langle
  x\rangle + \tau_1^2 \langle x^2 \rangle \right) \nonumber \\
&-&  \frac{2 e \gamma_+ \tau_0}{\hbar \tau_1} \sin \eta \langle p \rangle - e
\gamma_+ .
\end{eqnarray}
For $\eta \ne 0 \, {\rm mod} \, \pi$,
the average current contains a term proportional to the
average momentum of the oscillator that does not vary with the applied
bias \cite{wabnig2005}. 
However, since $\langle p \rangle =0$ in the
steady-state, the average current contains no information about the
momentum of the oscillator.
Therefore, the current of the detector can not be used
as a $p$-detector in the steady-state. Nevertheless, the current is important
to calibrate $\eta$ to the $p$-sensitive value $\pi/2$. A careful analysis of
the current $I$ as a function of $\Phi$ shows that the inflection points of
$I(\Phi)$ correspond precisely to values of $\eta= \pi/2 \, {\rm mod} \,
\pi$. Therefore, we can use a current measurement to tune $\eta$
to a $p$-sensitive value.

In the experimentally relevant regime, where $\tau_1^2 \langle x^2 \rangle \ll \tau_0^2$, and
for $\omega \sim \Omega$, the dominant contributions to the current
power spectrum of the detector are
\begin{eqnarray} \label{siom1}
&& S_I (\omega) = 2 e \langle I \rangle + 8 e^2 \omega   \int _0^\infty dt \,
\sin (\omega t) \times \nonumber \\
&& \left[ \frac{eV}{h} \cos \eta  \tau_0
    \tau_1\langle \langle x m \rangle \rangle  - \frac{\gamma_+
      \tau_0}{\hbar \tau_1} \sin \eta \langle \langle p m \rangle\rangle
  \right] \;,
\end{eqnarray}
where $\langle \langle ab \rangle \rangle = \langle a b \rangle -
\langle a \rangle \langle b \rangle$. We now further analyze the added
noise due to the presence of the oscillator, $\Delta S = S_I(\omega) -
2 e \langle I \rangle$. This noise spectrum is the sum of a
contribution arising due to correlations between the transfered charge
$m$ and position (term $\sim \langle \langle xm \rangle \rangle$ in
Eq. (\ref{siom1})), which we call $\Delta S_1$, and one due to
correlations between $m$ and the momentum of the oscillator (term
$\sim \langle \langle pm \rangle \rangle$ in Eq. (\ref{siom1})), which
we call $\Delta S_2$. The full spectrum is therefore $\Delta S =
\Delta S_1 + \Delta S_2$ with
\begin{eqnarray}
&&\Delta S_1 (\omega) = \Bigl[ \lambda_x^2 \left( 1 - \frac{\hbar \Omega}{2
    eV} \frac{\Delta x_0^2}{\langle \langle x^2 \rangle \rangle} \right)
\\
&&- \lambda_x \lambda_p \left ( \frac{M \Omega}{2 \pi}  \tau_1^2 \Delta x_0^2
  \frac{eV}{M \Omega^2 \langle \langle x^2 \rangle \rangle}  \right) \Bigr]
S_x (\omega) \nonumber \\
&&-  \lambda_x \lambda_p  \left(1 - \frac{M eV}{\langle \langle p^2 \rangle
    \rangle} \right)\frac{\langle \langle p^2 \rangle \rangle}{M} \frac{4  (\Omega^2-
  \omega^2) }{4 \gamma_{\mathrm{tot}}^2 \omega^2 + (\omega^2 - \Omega^2)^2} \nonumber \\
&&\Delta S_2 (\omega) = \lambda_p^2 \left( 1 - \frac{MeV}{\langle \langle p^2
    \rangle \rangle }\right) S_p (\omega) \\
&&+ \Bigl[ \lambda_p \lambda_x \left(
    1 - \frac{\hbar \Omega}{2 eV} \frac{\Delta x_0^2}{\langle \langle x^2
      \rangle \rangle} \right) \nonumber \\
&&-  \lambda_p^2 \left ( \frac{M \Omega}{2 \pi}
    \tau_1^2 \Delta x_0^2 \frac{eV}{M \Omega^2 \langle \langle x^2 \rangle
      \rangle}  \right) \Bigr] \frac{4 M \Omega^2 \langle \langle x^2 \rangle
  \rangle (\Omega^2- \omega^2) }{4 \gamma_{\mathrm{tot}}^2 \omega^2 + (\omega^2 -
  \Omega^2)^2} ,\nonumber
\end{eqnarray}
where the position and the momentum gain are given by
$\lambda_x = 2 e \tau_0 \tau_1 (eV/h) \cos \eta$ and $\lambda_p = (e/2 \pi M) \tau_0 \tau_1 \sin \eta$,
respectively. We now discuss several limits of the current
noise $S_I(\omega)$ of the detector in the case of a general phase $\eta$.
For $\eta=0 \, {\rm mod} \, \pi$, we recover Eq.~(30) of
Ref.~\cite{doiron2007} -- the position detector result. More interestingly,
for $\eta=\pi/2 \, {\rm mod} \, \pi$,
$\lambda_x = 0$ and the detector output contains only two
terms: The first one is proportional to $S_p (\omega)$ and therefore peaked
around $\Omega$. The second one is proportional to
$(\Omega^2 - \omega^2)$ and contributes
negligibly near resonance $\omega \approx \Omega$. Hence, for
$\eta=\pi/2$, we obtain
\begin{equation} \label{result}
\Delta S(\omega \approx \Omega) \approx
\lambda_p^2 \left( 1 - \frac{MeV}{\langle \langle p^2
    \rangle \rangle }\right) S_p (\omega) .
\end{equation}
Thus, the added noise is directly proportional to the momentum spectrum of the
oscillator. This is the key result of our Letter.

\begin{figure}[htb]
\includegraphics[width=8.5cm]{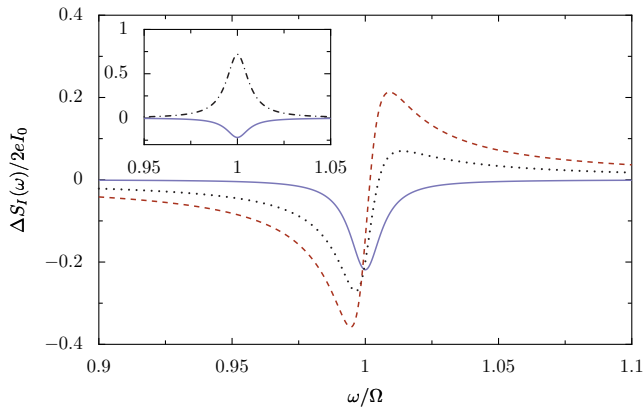}
\caption{(color online) Added current noise (normalized by $2eI_0 \equiv 2 e^3
  \tau_0^2 V/h$) of the proposed momentum detector due to the
  presence of the oscillator. For all curves,
  the bias is $eV=50 \hbar \Omega$, $\gamma_+ = \gamma_{\mathrm{tot}}/4$ and
  $\gamma_{\mathrm{tot}}=\Omega/200$. The main panel shows the total detector output
  for different values of the tunneling phase $\eta$ and for
  $T_{\mathrm{env}}=0$. The (blue) solid, (black) dotted, and (red)
  dashed lines correspond
  to $2 \eta / \pi=1,1.005$, and $1.01$, respectively.  In the inset, the
  current noise at the $p$-sensitive phase
$\eta=\pi/2$ is plotted for two different temperatures of the environment
 $T_{\mathrm{env}}=0$ (solid
  line) and $T_{\mathrm{env}}=5 eV/k_B$ (dash-dotted line).}
\label{fig_res}
\end{figure}

From the parameter dependence of each gain, we can estimate that the
momentum signal at $\eta=\pi/2$ should be typically smaller than the
position signal at $\eta=0$ by a factor $(eV / \hbar \Omega)^2$.
Nevertheless, it is unambiguously possible to identify a $p$ signal in
the current noise. We now describe three different ways to do
this. First, since $\lambda_x \propto V$ while $\lambda_p$ is
independent of $V$, the bias voltage dependence of the noise spectrum
can also be used to confirm that momentum fluctuations are
measured. Secondly, for an oscillator undergoing Brownian motion, the
temperature dependence of both signals differs qualitatively.
Like in the position detector case, the momentum signal is
reduced by a quantum correction (the term proportional to $- MeV /
\langle \langle p^2 \rangle \rangle$ in Eq.~(\ref{result})) that
arises from the finite commutator of $\hat{x}$ and $\hat{p}$. However,
there is a fundamental difference between the $x$-detector result
(Eq.~(7) of Ref.~\cite{clerk2004b}) and the $p$-detector result
(Eq.~(\ref{result})).  In the former case, the quantum corrections are
always small compared to the leading terms and therefore the peak
at resonance is always positive. In contrast, the two terms in
Eq.~(\ref{result}) can be of equal magnitude and compete about the
sign of $\Delta S(\omega \approx \Omega)$. The $p$-sensitive current
noise in Eq.~(\ref{result}) changes sign when the effective
temperature of the oscillator is equal to $(eV/k_B)(1-\gamma_+ /2
\gamma_{\mathrm{tot}}) /(1-\gamma_+ / \gamma_{\mathrm{tot}})$. For a
cold environment $T_{\mathrm{env}} \ll eV$, $\Delta S(\omega)$ is
negative at the resonance, whereas, for a hot environment
$T_{\mathrm{env}}> eV$, $\Delta S(\omega)$ is positive. This change of
sign never appears during a position measurement, so this pronounced
difference between a $x$-dependent and a $p$-dependent signal can be
used distinguish the two. We illustrate the change of sign in the
inset of Fig.~\ref{fig_res}, where the added current noise for
$\eta=\pi/2$ is plotted for different $T_{\mathrm{env}}$\cite{foot2}.

In the main panel of Fig. \ref{fig_res}, we plot the full detector
output for different values of $\eta$ near the optimal operation point
for momentum detection. Away from $\eta=\pi/2$, contributions to the
current noise $\sim \lambda_x^2$ become important and wash out the
momentum signal $\sim \lambda_p^2$. Indeed, for small $\Delta \eta =
\eta-\pi/2$, the ratio $\lambda_x / \lambda_p \sim (- \Delta \eta) (eV
/ \hbar \Omega)$ of the two amplification factors becomes large as
soon as $\lvert \Delta \eta \rvert > \hbar \Omega / eV$. In the
high-bias regime ($eV \gg \hbar\Omega$), momentum detection therefore
requires good experimental control over the applied flux. At moderate
bias $eV \geq \hbar \Omega$, the requirement on $\Delta \eta$ becomes
less restrictive. Finally, the current noise spectrum at $\eta=\pi/2$
shows a strong symmetry around $\Omega$ that makes the optimal
operation point easily identifiable.

In conclusion, we have shown how a modified tunnel junction
position detector can be designed to detect the
momentum fluctuations of a NEM oscillator. By using two tunnel
junctions in an AB-type setup, it is possible to precisely tailor
the interaction Hamiltonian between the detector and the
oscillator via an external magnetic field. We have demonstrated
how the proposed detector can be made sensitive to either
displacement or momentum fluctuations of the oscillator.

We would like to thank W.A. Coish, L.I. Glazman, O. G{\"u}hne, and I. Martin 
for interesting discussions. This
work was financially supported by the NSERC of Canada, the FQRNT of 
Qu{\'e}bec, the Swiss NSF, and the
NCCR Nanoscience.


\begin{thebibliography}{27}
\bibitem{foot3}
The overall
phase $\varphi_0$ is a gauge-dependent quantity and does not affect the
observable output of the detector.
\bibitem{foot5} Strictly speaking, the analysis in Eq.~(\ref{tau0p2}) 
is only valid in the single-channel case. For the multi-channel case with $N$
channels in the AB loop, the magnitude of the AB oscillations is reduced by a
factor $1/N$.
\bibitem{foot2}
The plots are done for an oscillator in the
thermal regime: $\langle \langle p^2 \rangle \rangle = M k_B
T_{\mathrm{eff}}$ with $\gamma_{\mathrm{tot}}T_{\mathrm{eff}} =
T_{\mathrm{env}} \gamma_0 + \gamma_+ (eV/2k_B)$.
\end{thebibliography}
\end{document}